# Using Microfluidic Device to Study Rheological Properties of Heavy Oil

Kiarash Keshmiri [a], Saeed Mozaffari [a, b], Plamen Tchoukov [a], Haibo Huang [c], Neda Nazemifard [a,*]

[a] Department of Chemical and Materials Engineering, University of Alberta, Edmonton, T6G 2V4 AB, Canada
[b] Department of Chemical Engineering, Virginia Polytechnic Institute and State University, Blacksburg, 24061 VA, United States
[c] Reservoir and Geosciences, Alberta Innovates Technology Futures

**Abstract**

In this study, capillary-driven flow of different pure liquids and diluted bitumen samples were studied using microfluidic channel (width of 30 µm and depth of 9 µm). Capillary filling kinetics of liquids as a function of time were evaluated and compared with theoretical predictions. For pure liquids including water, toluene, hexane, and methanol experimental results agreed well with theoretical predictions. However, for bitumen samples, as concentration of bitumen increased the deviation between theoretical and experimental results became larger. The higher deviation for high concentrations (i.e. above 30%) can be due to the difference between dynamic contact angle and bulk contact angle. Microchannels are suitable experimental devices to study the flow of heavy oil and bitumen in porous structure such as those of reservoirs

## 1. Introduction

Microfluidics is an emerging technology that deals with the fluid flow in micro-scale channels [1]. Microfluidic devices has been used as a good representative of porous media to evaluate dynamic aspects of fluid flow at pore-scale [2]. Micro and nanofluidic devices have received attention during the last few decades due to high control of hydrodynamics, fast and cheap fabrication, possibility of working at high temperature and pressure, and visualization of the fluid flow [3]. Prediction of capillary flow in micro and nanochannels are essential to develop emerging nanofluidic methods.

Capillary action drives fluid motion in microchannel without external driving forces. Capillary driven flow refers to spontaneous movement of interface due to curved liquid-liquid or gas-liquid interface [4]. In this study, glass etched microchannels with the depth of 10 µm and width of 30 µm were used to monitor the capillary-driven flow of different liquids. The size of channels is close to the size of real porous media of oil sands in Alberta, Canada. Capillary filling speed of hexane, methanol, toluene, and water as well as solutions of bitumen in hexane (10% to 50%) were experimentally monitored using inverted microscope with a digital charged coupled device (CCD) camera. In each chip, six parallel microchannels were fabricated which enables us to conduct several experiments with the similar experimental condition. This is important for repeatability of the data and having average values for each run.

Theoretical viscosity of each sample was calculated with the assumption of constant contact angle which was then compared with measured bulk viscosities. In our previous work in our group [5], capillary filling kinetics of bitumen solutions in nanochannel (depth ~ 47 nm) was investigated, where theoretical results were significantly deviated from experimental values specially for higher concentrations of bitumen samples. It seems that sharp variation of advancing contact angle and interface shape is the reason of this deviation (Figure 1). In fact, advancing contact angle is completely different from bulk contact angle and considering a constant value is not a reliable approach for high concentration of bitumen. Therefore, classical model failed to explain the filling kinetics of concentrated bitumen in the nanochannel.

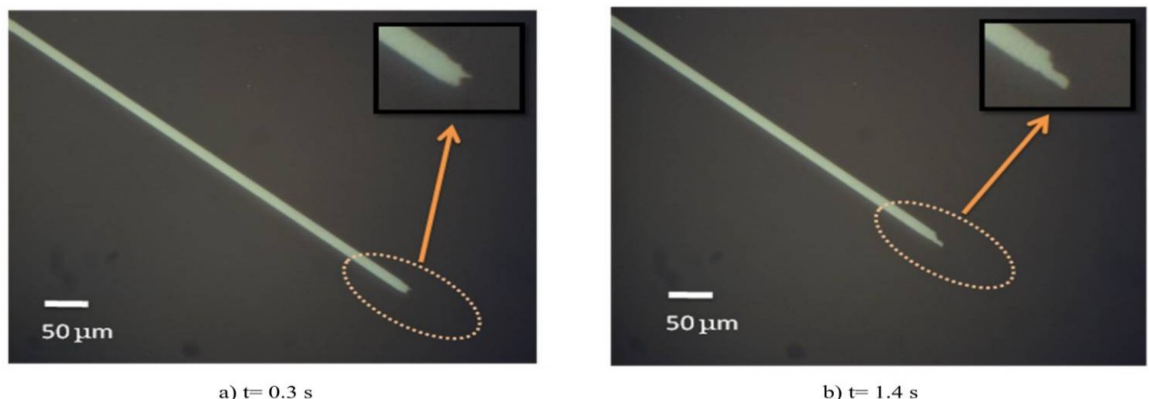

a) t= 0.3 s     b) t= 1.4 s

Figure 1. Capillary filling of 40% bitumen diluted in heptol (80:20) [5]

These findings further suggest that channel size can affect the wettability, where decreased wettability in nano-scale separations leads to large deviation between theoretical and experimental results. On the other hand, the experimental results of microchannels showed less deviation from classical model compared with nanoscale flows. It seems that microscale flow shows more predictable flow behavior compared with nanoscale using classical models while filling kinetics of viscous fluids in micro-scale require more study in order to have better prediction of real porous media structure.

## 2. Material and method

For capillary-driven flow study, methanol (HPLC grade, Fisher Scientific), hexane (HPLC grade, Fisher Scientific), toluene (HPLC grade Fisher Scientific) were supplied. Athabasca bitumen was also provided by Alberta Innovative Technology Feature (AITF) diluted with hexane (10% to 50%).

### 2.1. Microchannel fabrication

Photolithographic method was used for microchannel fabrication on 4 inch × 4 inch × 0.043 inch borofloat (81 % $SiO_2$, 13 % $B_2O_2$, 4 % $Na_2O/K_2O$, 2 % $Al_2O_3$) glass wafers. Wafers were cleaned at cleanroom using piranha solution (H2SO4 and H2O2 with 3:1 volume ratio) and rinsed to remove piranha solution using de-ionized water and then dried in spin rinse. A layer of Cr (75 nm) and Au (180 nm) was coated on the wafers as the masking layer. Photoresist (HPR 504) was spun onto the wafer and backed for 30 min. photoresist was exposed to light through a photomask that was previously designed using L-edit software in order to transfer the designed pattern to the photoresist. Then Au and Cr were wet etched with KI etchant based on photomask pattern. Borofloat glass was etched using hydrofluoric acid (HF) for the specific time based on desire depth of channel (1.4 µm/min). The remained Au and Cr and photoresist were removed from the wafer. Finally, the etched and unpatterned substrates were cleaned using piranha solution and pressed together to form temporary bonding. The advantage of temporary bonding is that it enables us to reuse the fabricated channels which significantly lowers the cost of fabrication. A general schematic of the fabrication process is illustrated in figure 2.

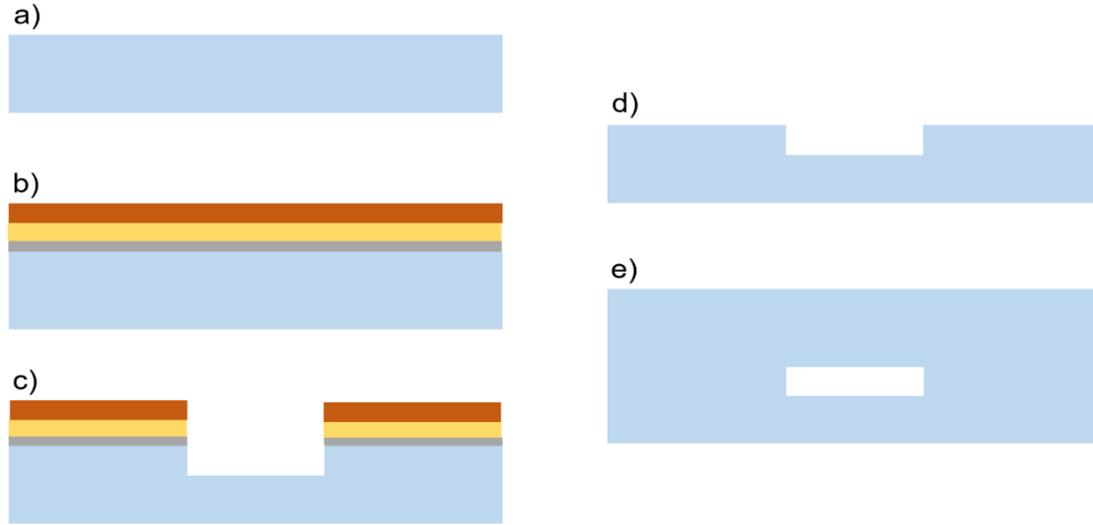

Figure 2. Microchannel fabrication process, a) Piranha cleaning, b) Metal layer and photoresist coating (Cr and Au), c) Developing of photoresist, etching the metals and borofloat, d) Removing the metal layers and photoresist, e) Glass-glass temporary bonding

**2.2. Viscosity, contact angle, and surface tension measurement**

Viscosity of diluted bitumen samples and other pure liquids was measured using rheometer (Rheolab QC, Anton Paar, Austria) with a double gap measurement system. Shear rate was ranged from 1 to 1000 $s^{-1}$. Measurement for each sample was repeated at least three times and temperature was maintained at $22 \pm 0.5$ °C. Bitumen samples showed Newtonain fluid behavior in agreement with our previous work which was done for slightly different condition [6]. Bulk contact angle of diluted bitumen samples and pure liquids was measured on the etched glass surface using FTA 200 equipped with camera and image analytical program. The same system (FTA 200) was used for air-liquids surface tension measurement using Young-Laplace equation to symmetric pendant drop. Bulk contact angles and surface tensions were measured three times at temperature of $22 \pm 0.5$ °C.

**3. Theory**

Experimental analysis of capillary filling kinetic of heavy oil in microchannels and further comparison with the theoretical models improves our understanding of fluid transport and wettability phenomena in the microscale. Fluid-fluid and fluid-surface interaction, geometry, wettability (oil-wet to water-wet), and surface tension are among the effective parameters in filling analysis. Capillary filling of liquids can be evaluated and compared with theoretical models such as Lucas-Washburn classical model [7]. This equation applies physical properties of the liquid and microchannel surface to predict the meniscus location as a function of time. Considering moderately constant contact angle and absence of gravity forces on the basis of Navier- Stokes law for an incompressible, laminar, and viscose flow the governing equation is written as [8-10]

$$-\frac{\partial P}{\partial x} = -\mu \frac{\partial^2 V_x}{\partial y^2} \qquad (1)$$

Where P is fluid pressure, x is penetration distance, μ the viscosity of the fluid, and $V_x$ fluid velocity. Pressure drop across the free surface of the meniscus is calculated based on Young-Laplace equation (Eq. 2) and combining with Eq. 1, leads to Lucas-Washburn equation (Eq. 3):

$$\Delta P = \frac{2\sigma \cos \theta}{r} \qquad (2)$$

$$x = \left(\frac{r\sigma \cos \theta}{2\mu} t\right)^{1/2} \qquad (3)$$

Where r is the hydraulic radius, σ the air-fluid surface tension, and θ the fluid advancing contact angle.

Average filling speed of advancing meniscus as a function of position can be written as (Eq. 4):

$$V_x = \frac{h\sigma \cos \theta}{4\mu x} \qquad (4)$$

## 4. Results and Discussion

Advancing liquid meniscus inside the microchannel was monitored using white light microscope coupled with high speed CCD camera. Recorded images were processed using ImageJ software. For all pure liquids and bitumen samples linear relation between square of position of advancing liquids and propagation time was found as expected in the case of Newtonian fluids. The experimental results are illustrated in figure 3 and compared with the Washburn model. It can be seen that theoretical prediction agrees well with experimental values.

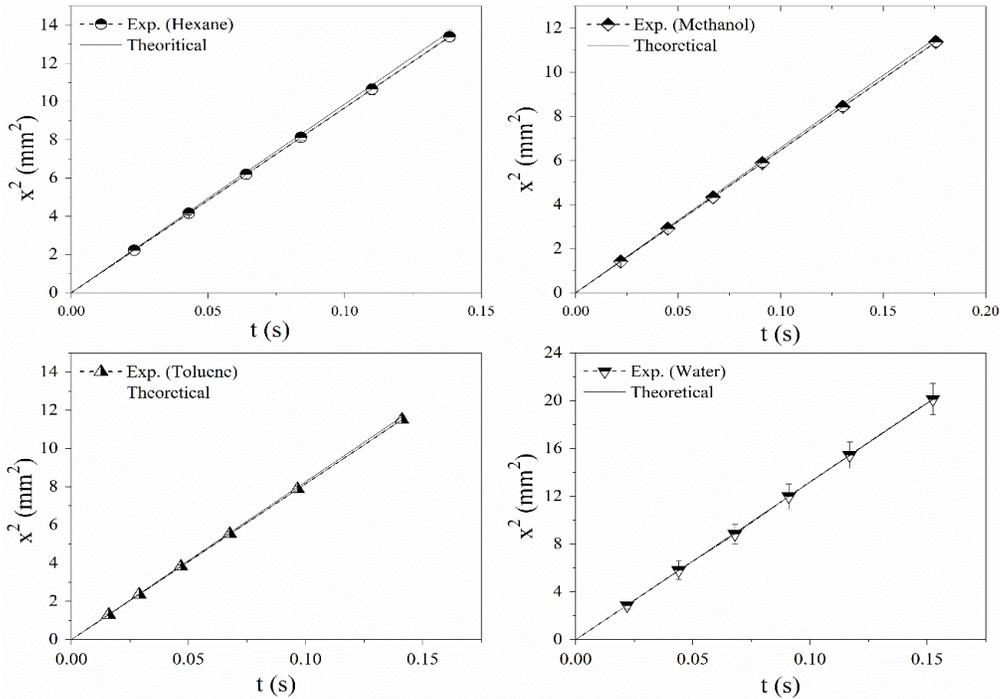

Figure 3. Experimental and theoretical comparison of pure fluid filling in microchannel

Averaged experimental data for the filling of water, methanol, toluene, and hexane in microscale channels were used to calculate fluid viscosity using the Washburn equation (Eq. 4). To determine the difference between theoretical and experimental values Average Absolute Relative Deviation (AARD%) is calculated using Eq. (6).

$$AARD\ (\%) = \frac{100}{N}\sum_{i=1}^{N}\left|\frac{x^{exp}-x^{cal}}{x^{exp}}\right| \qquad (6)$$

In this equation, N is the number of data points, $x^{exp}$ is the experimental position of the advancing interface and $x^{calc}$ is the calculated value obtained from Eq. 4. The summary of the results for pure liquids is shown in Table 1.

Table 1. Theoretical and experimental viscosity and AARD pure fluids

|  | Line slope $x^2/t$ (mm$^2$/s) | Theoretical viscosity (mPa.s) | Experimental viscosity (mPa.s) | AARD (%) |
|---|---|---|---|---|
| Water | 132.37 | 0.887 | 0.89 | 0.27 |
| Toluene | 81.58 | 0.567 | 0.56 | 0.64 |
| Hexane | 96.74 | 0.316 | 0.31 | 0.99 |
| Methanol | 64.67 | 0.596 | 0.58 | 0.8 |

Figure 4 shows time-lapsed images of capillary-driven flow of 40% bitumen in microchannel. The interesting observation during capillary flow in microscale is that the shape of interface is similar at different penetration distances as opposed to the continuous changes in the shape and contact angle of bitumen observed during capillary flow of bitumen in nanochannel [11]. It seems that effective forces exerted on the liquid in nanoscale are different from the micro scale. For instance, Phan et al [12] showed the importance of electrical double layer (EDL) in nano-scale, where the thickness of EDL is comparable to the length of channel and because of that the observed viscosity in the nanoscale was larger than the bulk value.

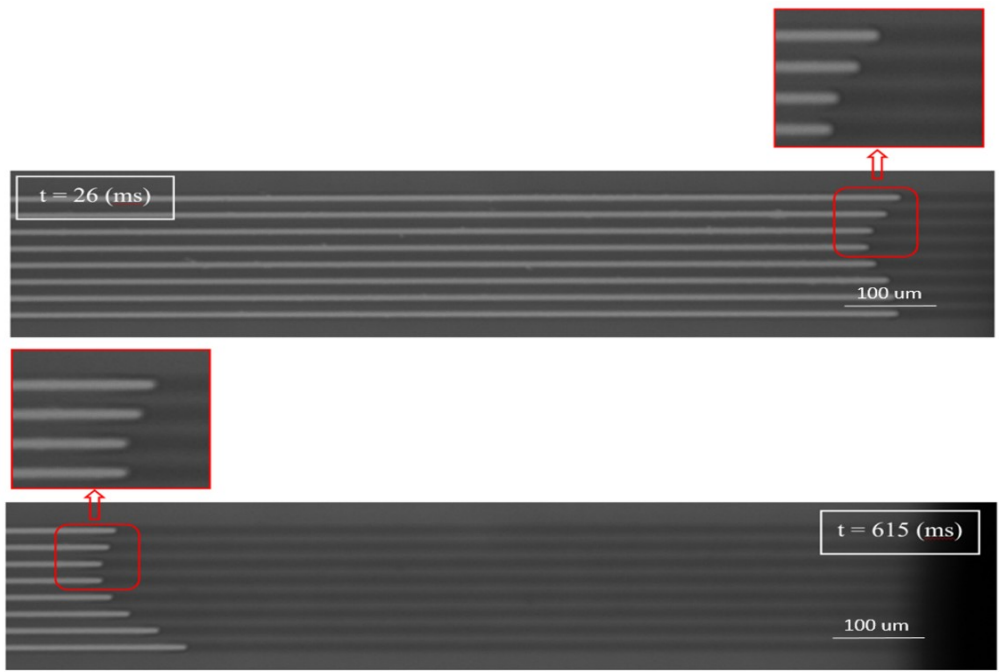
Figure 4. Time-lapsed images of capillary filling of diluted bitumen (40%)

In the case of diluted bitumen samples (10 wt.% to 50 wt.%), a linear relation between position squared of advancing liquids and propagation time was found (figure 5). In order to compare experimental results with theoretical predictions, $x^2/m$ as a function of time is plotted as shown in figure 6 ($m = r\sigma \cos\theta/2\mu$). Despite the good agreement between experimental and theoretical predictions of pure liquids, there was an increasing deviation when bitumen concentration increased from 10% to 50%. According to table 2, theoretical viscosities were calculated and compared with the experimental viscosities. Moreover, experimental and theoretical values for square meniscus position were compared based on AARD. This deviation between theory and experiment for higher bitumen fractions can be due to the change in the wettability and dynamic contact angle being different from the bulk.

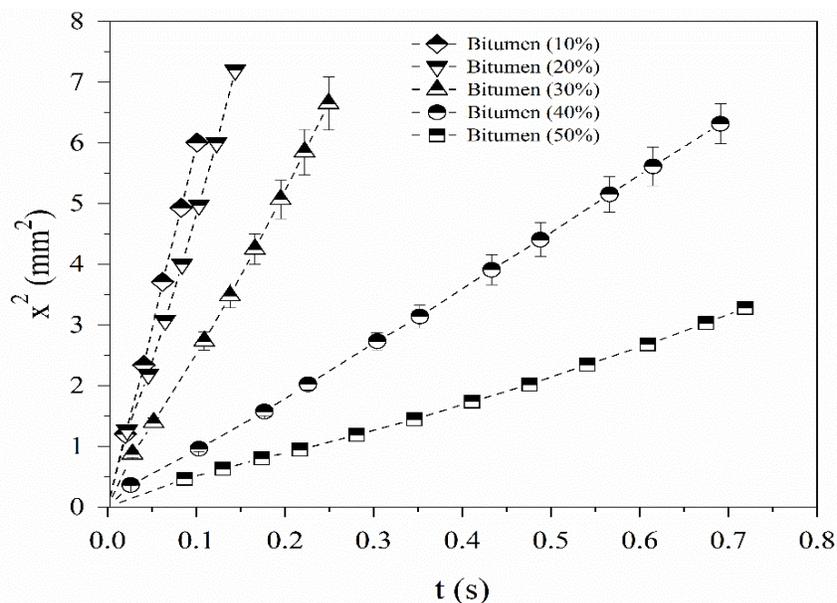
Figure 5. Square capillary flow of diluted bitumen vs time

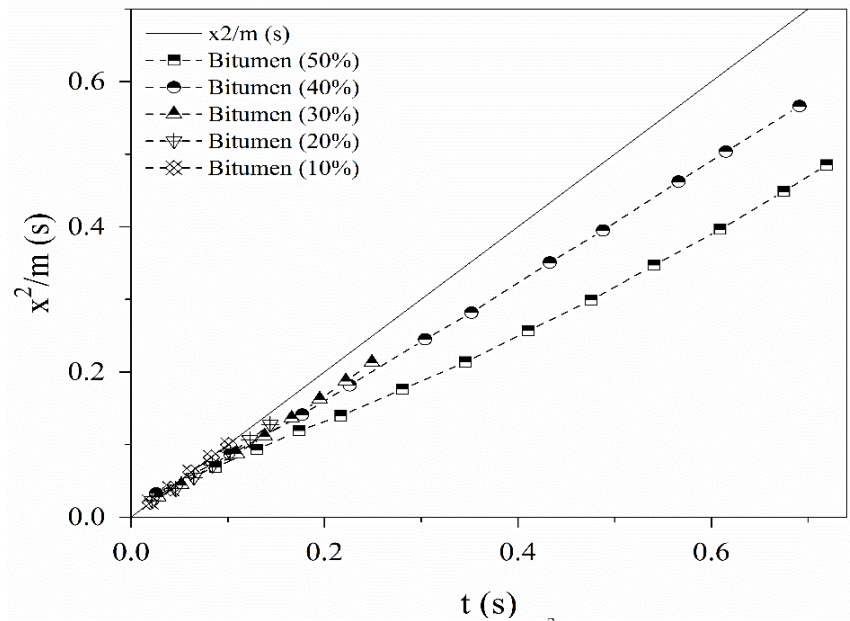
Figure 6. Diluted bitumen experimental data ($x^2/m$) as a function of t (s)

Table 2. Theoretical and experimental viscosity and AARD for diluted bitumen

|  | Line slope $x^2/t$ (mm$^2$/s) | Theoretical viscosity (mPa.s) | Experimental viscosity (mPa.s) | AARD (%) |
|---|---|---|---|---|
| Bitumen (10%) | 59.70 | 0.578 | 0.557 | 3.31 |
| Bitumen (20%) | 48.91 | 0.672 | 0.584 | 12.84 |
| Bitumen (30%) | 26.12 | 1.21 | 0.988 | 19.23 |
| Bitumen (40%) | 9.03 | 3.09 | 2.51 | 23.43 |
| Bitumen (50%) | 4.41 | 5.68 | 3.75 | 53.9 |

## 5. Conclusion

In this study, capillary filling of pure liquids and diluted bitumen (10%-50%) were studied and compared with theoretical predictions. For all the pure liquids and bitumen samples, a linear dependency between square position and time was observed which is an indication of Newtonian fluid type behavior. Capillary filling kinetics of pure liquids (i.e. methanol, toluene, hexane, and water) were in good agreement with the theoretical model (i.e. Washburn) with AARD less than 1%. For bitumen samples, theoretical predictions deviated from experimental results. This deviation was increased at higher concentrations of bitumen. It could be a result of surface wettability changes or difference between dynamic contact angle and bulk contact angle.